\documentclass[letterpaper, 10 pt, conference]{ieeeconf}  

\IEEEoverridecommandlockouts                              

\usepackage{graphicx} 
\usepackage{amsmath} 
\usepackage{amssymb}  
\usepackage{cases}
\usepackage{url}
\usepackage[backend=biber, style=numeric-comp, sorting=none, bibstyle=ieee]{biblatex} 
\bibliography{references.bib}

\title{\LARGE \bf
Speculative Thread Framework for Transient Management and Bumpless Transfer in Reconfigurable Digital Filters
}

\author{Michael Giardino$^{1}$, Wayne Maxwell$^{2}$, Bonnie Ferri$^{1}$ and Aldo Ferri$^{2}$
\thanks{$^{1}$Michael Giardino and Bonnie Ferri are with the School of Electrical and Computer Engineering,
        Georgia Institute of Technology
        {\tt\scriptsize giardino@gatech.edu bonnie.ferri@ece.gatech.edu}}%
\thanks{$^{2}$Wayne Maxwell and Aldo Ferri are with the School of Mechanical Engineering,
        Georgia Institute of Technology
        {\tt\scriptsize mrmaximus11@gatech.edu al.ferri@me.gatech.edu}}
}

\begin{document}

\maketitle
\thispagestyle{empty}
\pagestyle{empty}

\begin{abstract}

There are many methods developed to mitigate transients induced when abruptly changing dynamic algorithms such as those found in digital filters or controllers. 
These ``bumpless transfer" methods have a computational burden to them and take time to implement, causing a delay in the desired switching time. 
This paper develops a method that automatically reconfigures the computational resources in order to implement a transient management method without any delay in switching times. 
The method spawns a speculative thread when it predicts if a switch in algorithms is imminent so that the calculations are done prior to the switch being made. 
The software framework is described and experimental results are shown for a switching between filters in a filter bank.

\end{abstract}

\section{Introduction}

Common embedded processors have undergone dramatic improvements in recent years, both in processing power and in reconfigurable features.  
Low-level power management and temperature control features, for example,  allow for dynamic voltage and frequency scaling (DVFS) of the processor \cite{weiser_scheduling_1994,govil_comparing_1995} and changes in its sleep or idle state \cite{acpispecification}.   
In addition to DVFS and sleep state control, multicore processors have the ability to migrate loads across cores and turn on and off cores. 
Since the power consumption is directly proportional to clock frequency, voltage, and activity, adjusting these parameters can make a significant difference in power consumption.
For example, in the specific platform used in the paper, the Samsung Exynos SoC, there is a 3x increase in power consumption between 600 MHz and 2 GHz \cite{nikov_evaluation_2015}.

These features are either ignored or disabled when implementing a controller to ensure reliable and deterministic operation at the desired sampling rate. 
A strategy embraced in this paper is to design the control system for digital filter software to be ``compute-aware", that is, to monitor and take advantage of the dynamic reconfigurability of the processor when implementing digital filters or controllers. 
The result is a system that is designed with the flexibility of reconfiguration in both the controller or filter algorithm and in the processor.
Of particular interest in this paper is the implementation of transient management strategies in a compute-aware manner. 
Undesirable transients can occur in digital controllers or filters whenever the controller or filter is switched abruptly from one configuration to another. 
There has been much work done in the area of transient management in reconfigurable digital filter and control systems \cite{kovacshzy_transients_2001, valimaki_suppression_1998, piskorowski_dynamic_2007} 
Most of these papers consider the case when the structure of the filter or controller remains constant, only the parameters change. 

However, in the domain of digital filters and controls, drastic changes can occur due to the architecture of individual algorithms used. 
Namely, in situations where resources change sporadically, filters and controllers switch between algorithms of very different complexity. 
Transient management then becomes even more important, especially in those cases when switching from a low-order to high-order filter, since the initial conditions need to be chosen appropriately.

One approach to transient management in switched systems is termed ``bumpless transfer," which is a transient management strategy that aims to maintain a smooth output response in the presence of abrupt changes in the control or filter implementation; see, for example \cite{peng_antiwindup_1996}, \cite{qin_lq_2012}, and \cite{malloci_bumpless_2012}. 
In essence, the bumpless transfer technique uses various methods to compute the appropriate initial conditions for the new algorithm that would match the output of the old algorithms over some time interval. 
This eliminates ``bumps" in the output, but comes at the expense of computational effort, which grows higher as the filter order increases. 
This can result in delays and/or increased latency in the switching time. 
This paper proposes an alternate approach that is more amenable to real time applications.
	
A methodology called ``speculative threads" can be extended for improved implementation of bumpless transfer.  
Speculative threads are traditionally used in computing applications and refer to a method where serial tasks that may have data dependencies are computed speculatively based upon predicted data \cite{marcuello_speculative_1998,akkary_dynamic_1998,oplinger_enhancing_2002,martinez_speculative_2002}.
If the data were predicted correctly, then the work done by the thread is committed, and execution is accelerated.
If, on the other hand, the data were not predicted correctly, then the work is discarded and the execution proceeds as it would have without the speculation.
In this paper, the meaning of ``speculative threads" is expanded to describe a technique used to decrease latency in a processor by predicting when a thread might be needed and spawn it ahead of time in order to reduce the warm-up time. 

Speculative threads can be used as a means of implementing a predictive bumpless transfer method to reduce transients when switching dynamic algorithms. 
A computationally simple predication algorithm is developed to determine if a switch is imminent, and then it automatically spawns a thread that is used to perform transient mitigation strategies. 
At the same time, the computational resources are automatically increased so that there is no loss of performance on the running algorithm such as added latency.

The speculative thread development was combined with a means to manage the processor resources and the dynamic algorithm in a coordinated manner in order to come up with an overall compute-aware transient management strategy. 
In particular, CPU cores are idled or turned on by the digital filter application as computational load changes in an effort to maintain a constant latency (or time delay) of the main controller or digital filter algorithm. 
One solution that is possible with this general framework is to turn on an idle core or speed up a processor in order to compute the initial conditions for the bumpless transfer. 
While improving the transient performance, this method still results in some delay in the switching time of the filter. 
Another solution, based on the speculative thread concept, is to predict when a switch is imminent and then to turn on the idle core to run the new filter in parallel with the old one so that the switching transients die out sufficiently by the time that the new filter is switched in. 
The prediction algorithm is very simple and has very little computational load.

The general framework for using speculative threads for transient management methodology for switching dynamic algorithms is described in Section 2 while the practical means of implementing the method are described in Section 3. 
Experimental results are given in Section 4 for a digital filter.

\section{Speculative Threads for Transient Management}

As mentioned in the background section, improper transient management when a digital filter or controller reconfigures can have a dominant effect on the system performance \cite{ferri_reconfiguration}. 
A method that can be used to reduce induced transients is to run the new filter off-line and in parallel with the old one prior to the reconfiguration; that way, the new filter is in steady-state when it is switched on. 
There are two problems that must be overcome for this to be a viable strategy. 
First, it has a significant computational overhead, requiring two filters to be implemented concurrently. 
Second, waiting until the new filter reaches steady-state can cause a significant delay in the switching time, resulting in degraded performance. 
This paper shows how to solve these problems by developing a prediction scheme that anticipates when a switch is likely to happen, and launches a speculative thread to run the new filter if sufficient computational resources are available at that time.

\begin{figure}[!htbp]
	\centering
	\includegraphics[width=3in]{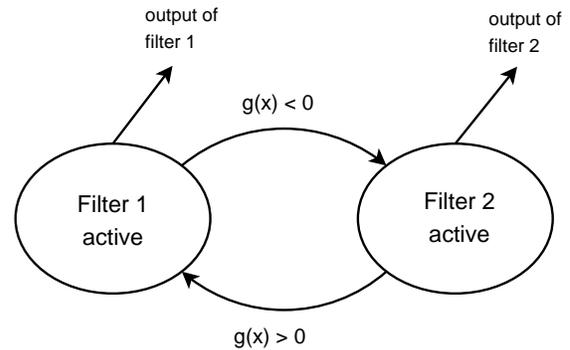}
	\caption{State machine for a reconfigurable controller with no transient management strategy.}
	\label{fig:threaded_state_machine}
\end{figure}

A simple, low-overhead predictor can be developed by modifying the existing switching condition calculations to include a component that determines a prediction if the switching boundary will be reached within a preset number of time steps. 
Then the algorithm is switched when it reaches that boundary with no delay, which may improve the response over other bumpless transfer methods that do incur a delay. 
The contributions in this area apply to any reconfigurable control system or digital filter. 
As the numbers of cores increases, using unutilized cores for speculative threads can reduce the performance penalties to the primary program and reduce the latency of switching the algorithm.

To describe the methodology, consider a digital filter with two possible configurations, Filter 1, $f_1(x,u)$, and Filter 2, $f_2(x,u)$, where x represents the states of the filter and u represents the filter inputs.  This system has a switching condition, $g(x)$, 
that mandates when the filter switches from one configuration to another:

\begin{equation} \label{eq:piecewise}
y =
\begin{cases} 
f_1(x,u)  & g(x) > 0 \\
f_2(x,u) & g(x) < 0
\end{cases}
\end{equation}

where $g(x)$ is the switching function. A state machine shown in Figure \ref{fig:threaded_state_machine} can be used to represent the state transitions for the simple switching implementation where no transient management strategy is used, that is, where the initial conditions for the new filter are chosen statically and \textit{a priori}.

\begin{figure}[!htbp]
	\centering
	\includegraphics[width=3in]{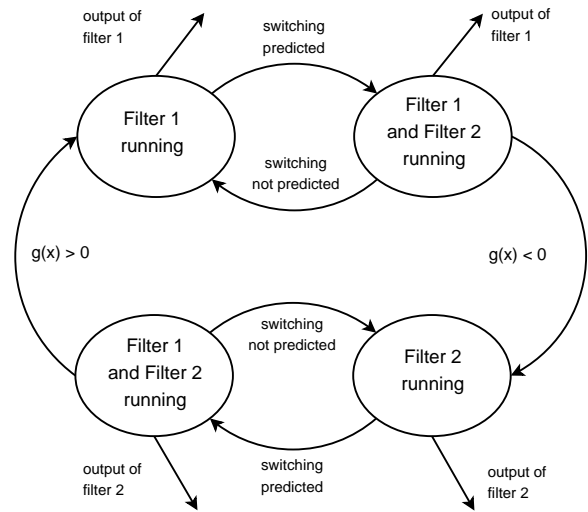}
	\caption{Example of a state diagram depicting two filter states and two prediction states.}
	\label{fig:spec_threaded_state_machine}
\end{figure}

To implement speculative threads, the state machine is enlarged with secondary switching conditions as shown in Figure \ref{fig:spec_threaded_state_machine}. 
If it is predicted that the switching surface will be breached in a small amount of time, then a speculative thread will be spawned to start implementing the secondary filter. 
That filter will become the primary filter when the switching surface is actually breached.

To determine the predication algorithm, a linear approximation is used to predict $g(x[n+m])$, $m$ steps into the future. 
For example, suppose $g(x[n]) < 0$ so that Filter 2 is running. 
The linear approximation for the value of the switching function m time steps in the future, $\tilde{g}(x[n+m])$ is given as

\begin{equation} \label{eq:switch_approx}
\tilde{g} = g(x[n]) + (g(x[n]) -g(x[n-1]) \cdot m
\end{equation}

where the finite difference $g(x[n]-g(x[n-1])$ is the instantaneous slope of the change in $g$. 
The approximation in equation \ref{eq:switch_approx} is used to predict if $g(x[n+m]) > 0$, that is, if the system would switch $m$ time steps in the future. 
Thus, the condition in equation \ref{eq:switch_approx} results in a simple prediction algorithm that is used to spawn a speculative thread to start implementing Filter 1, $f_1(x,u)$, on a new CPU core if

\begin{equation} \label{eq:switch}
g(x[n])(m+1) > g(x[n-1])m
\end{equation}

The value of $N=m$ is a design parameter selected based on the time constant of the filter. 
If the controller or digital filter has a time constant of $M$ steps, then $N$ might be chosen as $N=M$ or $N=2M$ in order for the transient to decay sufficiently before the switch is made. 
Note that $g(x[n])$ is being calculated anyway to determine when to switch controllers, so the incremental calculations needed for prediction are very minor. 
This design parameter might be called a prediction horizon since it relates to how far in advance of a switch that the prediction is made.

To summarize the secondary switching conditions that pertain to the prediction shown in Figure \ref{fig:spec_threaded_state_machine}: If filter 1 is running, that is, $g(x) > 0$ then switching is predicted if $g(x[n])(N+1) < g(x[n-1])\cdot N$.
If filter 2 is running (i.e $g(x) < 0$) then switching is predicted if $g(x[n](N+1) > g(x[n-1])\cdot N$.

\section{Generic Implementation Framework}

The goal of this research is to create a portable library for speculative multi-threading that can be used on any symmetric multi-programmable system running Linux. 
Our library was developed entirely in C, with the intent of making it fast, modifiable, and portable across any OS that can use POSIX threads (pthreads).
By leveraging pthreads we were able to build speculative threads on a robust base, allowing other researchers and software engineers to easily use and modify the framework. 

\begin{figure}[!htbp]
	\centering
	\includegraphics[width=2.25in]{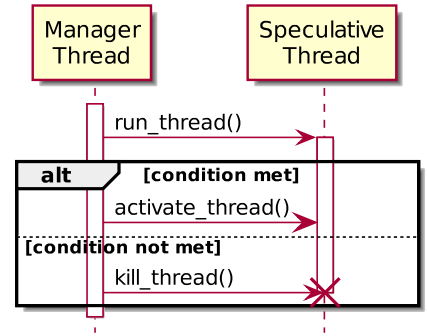}
	\caption{UML sequencing diagram for a generic speculative thread.}
	\label{fig:generic_spec_thread}
\end{figure}

The software implementation using speculative threads can be illustrated using a Unified Modeling Language (UML) sequencing diagram. 
UML is a graphical representation for designing and analyzing software systems \cite{rumbaugh_uml_2004}.
The sequencing diagram is especially useful for depicting the interaction of multiple processes (or threads) including data dependencies, external events and inputs, and the sequencing of each process.
Figure \ref{fig:generic_spec_thread} shows a sequencing diagram for a typical speculative thread used in general non real-time applications. 
The diagram shows two threads, a manager that controls the other threads, including the speculative thread. 
The threads are shown being created at the top of the diagram, then time runs downward showing the sequence of events such as when each thread is running and what events cause a change in the threads.
The thin rectangular blocks below each thread indicate when that particular thread is running; the manager thread is always running. 
The speculative thread is spawned when it is predicted it might be needed.
In the figure, this is indicted by the \texttt{\small run\_thread()} command.
The box labeled "alt" shows the alternate paths that a branch (or if-then) decision might take; one branch is above the dashed line and the other is below it. 
For the first branch, the condition is met that the speculative branch is no longer speculative, that is, it is needed in the operation and an \texttt{\small activate\_thread()} command is used.
The other branch is that the condition is not met for that speculative branch to continue, so it is terminated with the \texttt{\small kill\_thread()} command.

\begin{figure}[!htbp]
	\centering
	\includegraphics[width=3.25in]{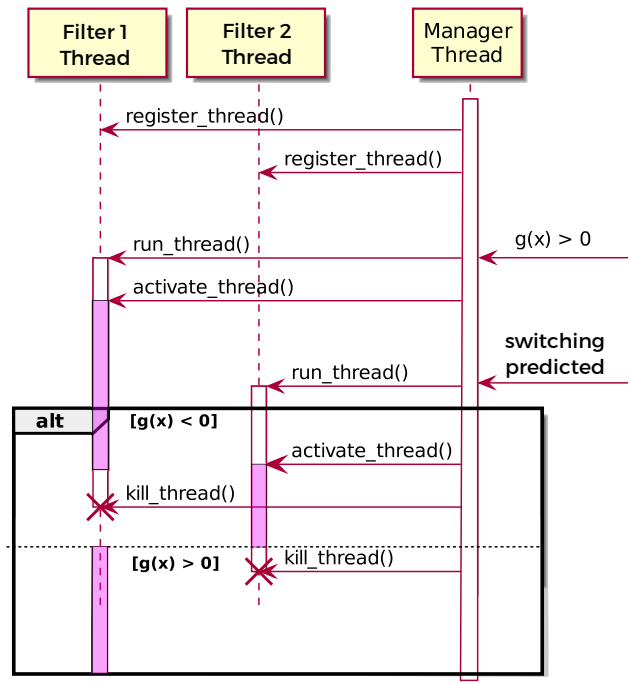}
	\caption{UML sequencing diagram for a speculative threads to implement transient management.}
	\label{fig:filter_spec_thread}
\end{figure}

Figure \ref{fig:filter_spec_thread} shows a UML sequencing diagram for the speculative threads that implement the digital filter system transient mitigation state diagram in Figure \ref{fig:spec_threaded_state_machine}. 
There are three threads that can run simultaneously: the Manager thread, Filter 1 (computes $f_1(x,u)$), and Filter 2 (computes $f_2(x,u)$). 
The Manager thread not only computes $g(x)$ and the switching conditions for the prediction, it controls when to run the other threads and when to reduce or increase computational resources. 
The computational resources are increased when both threads are running so that the execution time does not change for the primary filter, the one that is outputting the filter response. 
Resources are controlled by changing the frequency of the processor or by starting up or shutting down additional cores on a multicore processor.

The diagram in Figure \ref{fig:filter_spec_thread} can be explained by describing the various parts. Again, the sequence of events goes downward.  In the initial state, Filter 1 is running since $g(x)>0$.  
When the prediction is made, the Filter 2 thread is started with the \texttt{\small run\_thread()} command. 
Both threads continue until a branch is reached where either the primary switching
condition is met, $g(x)<0$ indicating that the Filter 1 thread should be terminated, or that the prediction is no longer valid and that the primary switching condition has not been met  so that the Filter 2 thread should be terminated.  
In the figure, the ``alt" block shows the decision logic between two possible ways to exit the prediction state; these alternative branches are both shown but are separated by a dashed line. 
The primary switching and prediction switching conditions used to change states in Figure \ref{fig:spec_threaded_state_machine} are shown next to the Manager thread in Figure \ref{fig:filter_spec_thread}. 
The corresponding actions are to run or stop threads and to change the computer resources.

\section{Application to Reconfigurable Filter Banks}

The speculative thread method to reduce transients was tested on a filter bank consisting of two filters: a third-order and a second-order Chebyshev low-pass filters.  The filter bank and the speculative 
thread framework were implemented on a ODROID-XU4 single-board computer, which contains ARMs big.LITTLE heterogeneous processor containing eight cores (4 Cortex-A15 and 4 Cortex-A7).
The big.LITTLE architecture is an attractive platform because of its mainline Linux support for scheduling threads on both high- and low-power cores contained on the same die \cite{amd_big_little}.


A decision variable is needed to determine which filter to use during a hybrid operation.  
This decision variable might represent an external sensor signal or an internal signal related to the available computational resources.
For example, when resources are high then a more computationally intense filter, such as the higher-order filter, might be used. When
resources are scarce, then a less costly low-order filter might be used. For repeatability, the decision variable was defined for this experiment as a
low-frequency sinusoid representing a time-varying computational load that causes a sinusoidal variation in the computational resources $\frac{1}{2}(1+sin(t))$. 
When the simulated load was less than 50\% (ie, the sinusoidal value was less than 0.5), the high-order filter was run, and when the load was greater than 50\%, the low order filter was run. The input to the filter was a higher frequency sine wave.

\subsection{Speculative Threading}

\begin{figure}[h]
	\includegraphics[width=3.25in]{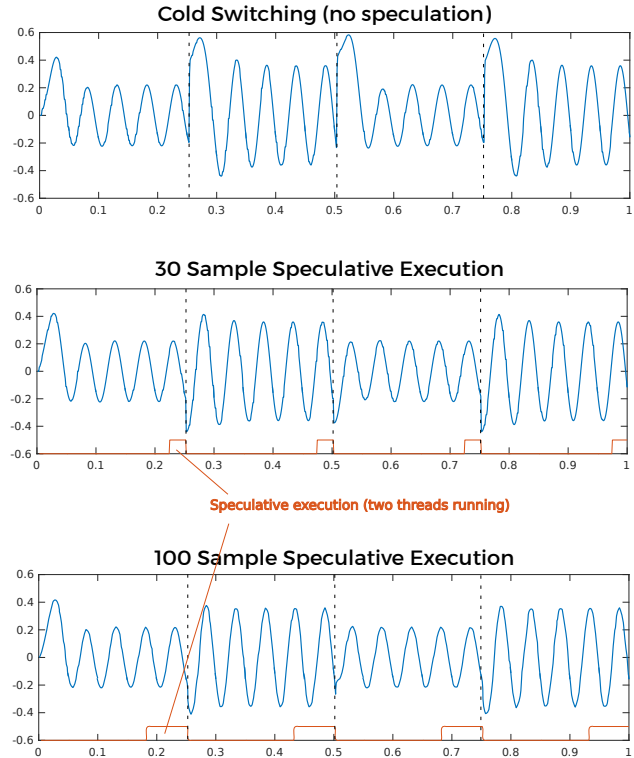}
	\caption{Experimental results of the output of the filters versus time for cold switching and for two cases using speculative threads, with 30 sample and 100 sample prediction horizons.}
	\label{fig:filter_cold_30_100}
\end{figure}

\begin{figure*}[!htpb]
	\centering
	\includegraphics[width=7in]{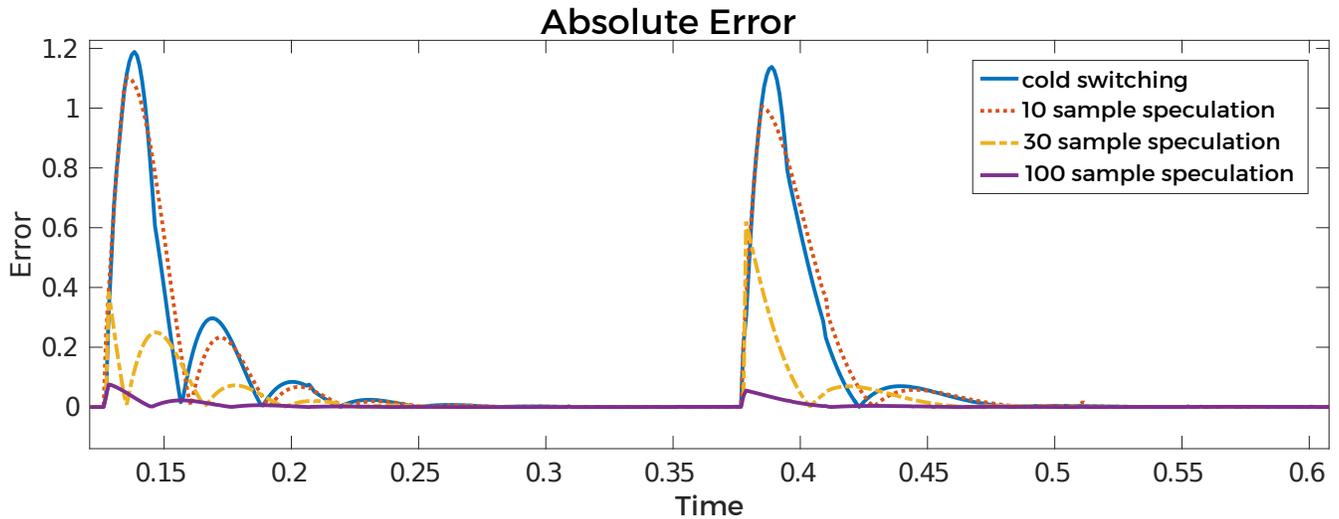}
	\caption{Comparison of each switching case to the benchmark always-on filter bank case.}
	\label{fig:rms_plot}
\end{figure*}

We examined three different scenarios of switching between the two filters, where they all used the same decision logic based on a sinusoidally varying load described above.
In the first scenario, the benchmark for comparison, both filters were run constantly and multiplexed to a single output according to decision logic. 
There are no transients due to initializing the filters, but then the computational resources are doubled.
In the second case, the filters were switched cold, that is, initialized at each switch with zeroed states and no time for the transients to decay. 
In the final scenario, the speculative thread method was used. 
The filter threads were launched when the computational load was predicted to pass the 50\% threshold. 
Comparison of each switching case to the benchmark always-on filter bank case shows significant improvement as the amount of warm-up time increases.

The experimental results are shown in Figure \ref{fig:filter_cold_30_100} for cold switching and for two speculative thread cases. 
The vertical dashed lines indicate the times when the filters are switched from 2nd to 3rd order (or vice versa).
The cold switching case switches the filter by setting the initial conditions for the new filters to be zero. 
The speculative thread cases spawn the new filter thread 30 or 100 samples prior to the predicted switching time and then switch to the new thread when the switching condition is actually reached. 
An indication of the time spent during speculation execution (when both filter algorithms are being computed) is shown at the bottom of the speculative thread plots. 
The benchmark case, with no switching transients, would show nearly pure sinusoids from about 0.1 seconds onward, with only a change in amplitude and phase due to the switching of filters. 
Of the three cases shown in the figure, the 100-sample speculative thread case is closest to the ideal benchmark case. 
The time constants for both of these filters are approximately 30 samples. 
Thus, the 30-sample prediction case allows the filters to run for one-time constant prior to switching, while the 100-sample prediction case allows the transients to decay for approximately three time constants prior to switching in the filters.  


For a quantitative examination of the technique, Figure \ref{fig:rms_plot} plots the error magnitude, that is the difference between the speculative and cold switching scenarios and the benchmark case.
The performance of speculatively launched filters improves significantly over the cold switched variant, especially as the length of the warm-up period increases.

Cold switching has a maximum error of 1.189 and an RMS error of 0.2801.
Adding 10-cycle speculation only reduces the maximum and RMS error by 5.63\% and 5.82\% respectively.
A 30-sample speculation improves maximum and RMS error by 47.86\% and 66.22\%.
With a 30-sample speculation, two filters run simultaneously only approximately 6\% (60 of 1000 samples) of the total execution time but it shows a tremendous improvement.
If we extend speculation to 100 samples (approximately 20\% execution overlap), we obtained a 93.52\% reduction in maximum error and 95.72\% decrease in RMS error.

\subsection{Introducting Hysterisis}

\begin{figure}[!htpb]
	\includegraphics[width=3.5in]{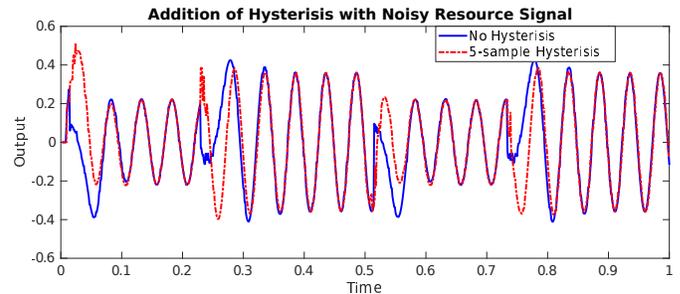}
	\caption{Influence of the hysteresis on switching transients with a noisy resource/switching signal.}
	\label{fig:deathwatch}
\end{figure}

The above scenarios were run with a resource/switching signal that is a pure sine wave, while a noisy or irregular switching signal offers different challenges. 
For example, a noisy signal might oscillate across a switching surface, that is, trigger the switching condition repeatedly in a short amount of time. In switched systems, this is known as chattering.  In this situation, the prediction condition as well as the switching condition might chatter. In terms of speculative threads, chattering might cause the speculative thread to be alternatively killed and then reactivated repeatedly. When it is reactivated, the filter in the speculative thread is set to zero initial conditions.
To alleviate this chattering, we introduce a time-based hysteresis to stop the premature termination of a speculative thread by keeping a deactivated thread alive for a specified amount of time. 
Experimental results suggest a hysteresis delay of 10\% of the speculation period, that is, the value of N, is a good rule of thumb. 

Figure \ref{fig:deathwatch} shows an example of the problems introduced by a noisy switching signal. 
Gaussian noise ($\mu=0.1$ and $\sigma=0.5$) was added to the switching sinusoidal signal. 
Without the hysteresis mechanism, the speculative filter threads are killed immediately when the prediction is no longer valid. 
As seen in Figure \ref{fig:deathwatch}, by adding time-based hysteresis, the transients are mitigated even in the presence of a noisy switching signal.   

\section{Conclusions and Future Work}

Speculative threads is a promising transient management implementation method that employs reconfigurable computing mechanisms available on many embedded system platforms.
The speculative thread framework developed in this paper handles transient mitigation for switching between dynamic algorithms such as digital filters or digital controllers. 
The prediction algorithm that triggers a speculative thread to be started is very simple and has little computational overhead. 
The fact that the algorithm is predictive means that the computations are done prior to the switching time resulting in smaller latency in the switching time compared to traditional transient management strategies that are implemented after the switch condition is met. 
Also, the compute-aware nature of the developed framework allows for the control of the processor resources such that the processor speed may be increased or idle cores started in order to run the speculative thread in parallel with no loss in resources or performance for the active filter thread.   

Experimental results run on an ODROID platform demonstrates the feasibility and benefits of the speculative thread framework. 
A design parameter, $N$, called the prediction horizon was introduced that dictates when a speculative thread is triggered to start. 
The larger the value of $N$, the longer the the speculative thread runs prior to a switch.
Correspondingly, the larger the value of $N$, the smaller the switching transients in the experimental results. 
A hysteresis mechanism introduced to avoid chattering keeps speculative threads alive briefly after the prediction of an impending switch ends in case the prediction is retriggered in a short period. 
This hysteresis method improves the performance of the speculative threads in environments when the switching signal is noisy.  

Speculative threads were demonstrated on linear digital filters, but it is equally applicable to other dynamic alogithms. 
We are currently doing experiments in a closed-loop hybrid motor controller that employs speculative threads for transient management.





\section*{ACKNOWLEDGMENT}

This work was funded via NSF CMMI 1538877.


\printbibliography

\end{document}